\documentclass[conference]{IEEEtran}

\IEEEoverridecommandlockouts
\usepackage{amsfonts}
\usepackage{amsmath}
\usepackage{balance}
\usepackage{multicol}

\usepackage{hyperref}

\usepackage{stfloats}

\usepackage[dvips]{graphicx}
\begin{document}

\hyphenation{multi-symbol}

\title{Adjacent-Channel Interference in Frequency-Hopping Ad Hoc Networks\vspace{-0.2cm}}
\author{\IEEEauthorblockN{Matthew C. Valenti,\IEEEauthorrefmark{1}
Don Torrieri,\IEEEauthorrefmark{2}
and Salvatore Talarico\IEEEauthorrefmark{1} }
\IEEEauthorblockA{\IEEEauthorrefmark{1}West Virginia University, Morgantown, WV, USA. \\
\IEEEauthorrefmark{2}U.S. Army Research Laboratory, Adelphi, MD, USA.}
\vspace{-0.5cm}
\thanks{M.C. Valenti and S. Talarico were sponsored by the National Science Foundation under Award No. CNS-0750821.}
}
\date{}
\maketitle

\vspace{0.5cm}
\thispagestyle{empty}

\begin{abstract}
This paper considers ad hoc networks that use the combination of coded continuous-phase frequency-shift keying (CPFSK) and frequency-hopping multiple access.  Although CPFSK has a compact spectrum, some of the signal power inevitably splatters into adjacent frequency channels, thereby causing adjacent-channel interference (ACI).  The amount of ACI is controlled by setting the fractional in-band power; i.e., the fraction of the signal power that lies within the band of each frequency channel.    While this quantity is often selected arbitrarily, a tradeoff is involved in the choice.     This paper presents a new analysis of frequency-hopping ad hoc networks that carefully incorporates the effect of ACI.  The analysis accounts for the shadowing, Nakagami fading, CPFSK modulation index, code rate, number of frequency channels, fractional in-band power, and spatial distribution of the interfering mobiles.  Expressions are presented for both outage probability and transmission capacity.    With the objective of maximizing the transmission capacity, the optimal fractional in-band power that should be contained in each frequency channel is identified.
\end{abstract}

\section{Introduction} \label{Section:Intro}
The combination of coded continuous-phase frequency-shift keying (CPFSK) and frequency-hopping (FH) spread spectrum is an attractive choice for ad hoc networks.   Compared with direct-sequence spread spectrum, FH can be implemented over a much larger frequency band and does not require power control to prevent the near-far problem.   The performance of FH systems depends upon the number of available frequency channels.  Increasing the number of frequency channels decreases the probability of collision, though this comes at the expense of requiring narrower signal bandwidths, which generally reduce the transmission rate.

For FH systems, continuous-phase frequency-shift keying (CPFSK) is the preferred modulation.   Frequency hopping with CPFSK offers a constant-envelope signal, a compact signal spectrum, a robustness against both partial-band and multiple-access interference, and is suitable for noncoherent reception \cite{torrieri:2011}.   CPFSK (assumed here to be binary) is characterized by its modulation index $h$, which is the normalized tone spacing.    Because the bandwidth is proportional to $h$, a decrease in $h$ increases the number of available frequency channels and improves the resistance to multiple-access interference.  However, the energy-efficiency of CPFSK generally improves with increasing $h$.
When the ad hoc network uses channel-coded CPFSK, the performance is also a function of the rate of the error-control code.   In \cite{valenti:2012}, the tradeoff among modulation index, code rate, and number of frequency channels was explored, and the values were optimized with the objective of maximizing the (modulation-constrained) {\em transmission capacity} \cite{weber:2005}, which is a measure of the spatial spectral efficiency.

% (assumed here to be constant)

The spectral efficiency of CPFSK is quantified by its {\em X-percent-power bandwidth}, which is the bandwidth containing $X$ percent of the power.  In \cite{valenti:2012}, the frequency channels were assumed to be separated by the 99-percent power bandwidth; equivalently, the bandwidths of the frequency channels were matched to the 99-percent power bandwidth of the modulation.  This implies that one percent of the signal power splatters into adjacent bands, but in \cite{valenti:2012} the adjacent-channel interference (ACI) due to spectral splatter was neglected.  However, the ACI can cause a non-negligible performance degradation, which needs to be taken into account by the analysis and optimization.

Matching the bandwidth of the frequency channels to the 99-percent power bandwidth of the modulation is a typical, but arbitrary, choice and is motivated by the desire to neglect the ACI.   It is possible to increase the bit rate while fixing the channel bandwidth, thereby decreasing the fractional in-band power and increasing the fraction of the signal power that splatters into adjacent channels.    While the resulting increased ACI will negatively affect performance, the increased bit rate can be used to support a lower-rate error-correction code, thereby improving performance.     Thus, there is a fundamental tradeoff involved in determining the fractional in-band power that should be contained within each frequency channel.  Quantifying this tradeoff requires analysis of the effects of spectral splatter.

The contributions of this paper are as follows.  The paper presents an analysis of FH systems that accurately accounts for the effects of ACI in the presence of shadowing and Nakagami fading. Expressions are given for the conditional outage probability, where the conditioning is with respect to the location of the mobiles and the realization of the shadowing, as well as for the spatially averaged outage probability and transmission capacity.  Having carefully modeled the effects of ACI, the paper refines the optimization of \cite{valenti:2012} by identifying the appropriate amount of spectral splatter, or equivalently the optimal fractional in-band power that should be contained in each  frequency channel.

\section{Network Model} \label{Section:SystemModel}
The network comprises $M+2$ mobiles that include a reference receiver located at the origin, a source or reference transmitter $X_{0}$, and $M$ interfering mobiles $X_{1},...,X_{M}.$   The variable $X_{i}$ represents both the $i^{th}$ mobile and its location, so that $|X_{i}|$ is the distance from $X_i$ to the receiver.   It is assumed that the interfering mobiles are constrained to lie in an annulus with inner radius $r_{\mathsf{ex}}$, outer radius $r_{\mathsf{net}}$, and area $A=\pi(r_{\mathsf{net}}^2-r_{\mathsf{ex}}^2)$.  A nonzero $r_{\mathsf{ex}}$ may be used to model an exclusion zone  \cite{torrieri:2012a}, which represents a minimum distance to a potential interferer that is imposed by the multiple-access protocol or constraints on the placement of the mobiles.

At the reference receiver, $X_i$'s  power is
\begin{eqnarray}
  \rho_i
  & = &
  P_i g_i d_0^\alpha 10^{\xi_i/10} |X_i|^{-\alpha} \label{eqn:power}
\end{eqnarray}
where $d_0$ is a reference distance, $P_{i}$ is $X_i$'s transmitted power measured at the reference distance $d_0$, $g_i$ is the power gain due to fading, $\xi_i$ is a shadowing coefficient, and $\alpha > 2$ is the attenuation power-law exponent.  Each $g_i = a_i^2$, where $a_i$ has a Nakagami distribution with parameter $m_i$, and $\mathbb E[g_i] = 1$.  For Rayleigh fading, $m_i=1$.

%In the presence of log-normal shadowing, the $\{ \xi_i \}$ are independent and identically distributed zero-mean Gaussian with standard deviation $\sigma_\mathsf{s}$ dB.  In the absence of shadowing, $\xi_i = 0$.

Channel access is through a slow frequency-hopping protocol.  It is assumed that the \{$g_{i}\}$ remain fixed for the duration of a hop, but vary independently from hop to hop (block fading).  While the $\{g_{i}\}$ are independent, the channel from each transmitting mobile to the reference receiver can have a distinct Nakagami parameter $m_i$.  An overall spectral band of $B$ Hz is divided into $L$ contiguous frequency channels, each of bandwidth $B/L$ Hz.  The mobiles independently select their transmit frequencies with equal probability.  The source $X_0$ selects a frequency channel at the edge of the band with probability $2/L$ and a frequency channel not at the edge with probability $(L-2)/L$.  Let $D_i \leq 1$ be the duty factor of $X_i$, which is the probability that the mobile transmits any signal.

Two types of collisions are possible, {\em co-channel} collisions, which involve the source and interfering mobile selecting the same frequency channel, and {\em adjacent-channel} collisions, which involve the source and interfering mobile selecting adjacent channels.
Let $p_\mathsf{c}$ and $p_\mathsf{a}$ denote the probability of a co-channel collision and an adjacent-channel collision, respectively.  Assume that $D_i = D$ is constant for every mobile.
%First consider co-channel collisions.
If $X_i$, $i\geq 0$, transmits a signal, then it will select the same frequency as $X_0$ with probability $1/L$.  Since $X_i$ transmits with probability $D$, the probability that it induces a co-channel collision is $p_\mathsf{c} = D/L$.
% Now consider adjacent-channel collisions.
If $X_i$, $i\geq 0$, transmits a signal, it will select a frequency channel adjacent to the one selected by $X_0$ with probability $1/L$ if $X_0$ selected a frequency channel at the edge of the band (in which case, there is only one adjacent channel), otherwise it will select an adjacent channel with probability $2/L$ (since there will be two adjacent channels).  It follows that, for a randomly chosen channel, the probability that $X_i$,  $i\geq 0$, induces an adjacent-channel collision is
\begin{eqnarray}
p_\mathsf{a} & = & D\left[ \left( \frac{2}{L} \right) \left( \frac{L-2}{L} \right) + \left( \frac{1}{L} \right) \left( \frac{2}{L} \right) \right] = \frac{2D(L-1)}{L^2}. \nonumber \label{eqn:p_a}\\
\end{eqnarray}

Only a certain percentage of a mobile's transmitted power lies within its selected frequency channel.  Let $\psi$ represent the {\em fractional in-band power}, which is the fraction of power in the occupied frequency channel.  Typically $0.95 \leq \psi \leq 0.99$, though a goal of this paper is to determine the optimal value of $\psi$.
% It follows that $\bar{\psi} = 1 - \psi$ is the {\em fractional out-of-band power}; i.e., the fraction of power outside the occupied frequency channel.
We assume that a fraction $K_s = (1-\psi)/2$, called the {\em adjacent-channel splatter ratio}, spills into each of the frequency channels that are adjacent to the one selected by a mobile.  For most practical systems, this is a reasonable assumption, as the fraction of power that spills into frequency channels beyond the adjacent ones is negligible.

Under the model described above, the instantaneous signal-to-interference-and-noise ratio (SINR) at the receiver is
\begin{eqnarray}
   \gamma
   & = &
   \frac{ \psi \rho_0 }{ \displaystyle {\mathcal N} + \sum_{i=1}^{M} I_i \rho_i } \label{Equation:SINR1}
\end{eqnarray}
where $\mathcal N$ is the noise power and $I_i$ is a discrete random variable that may take on three values with the probabilities
\begin{eqnarray}
I_i
& = &
\begin{cases}
  \psi & \mbox{with probability $p_\mathsf{c}$} \\
  K_s & \mbox{with probability $p_\mathsf{a}$} \\
  0 & \mbox{ with probability $p_\mathsf{n}$ }
\end{cases}
\label{eqn:I_i}
\end{eqnarray}
where $p_\mathsf{n} = 1 - p_\mathsf{c} - p_\mathsf{a} = 1-D(3L-2)/L^2$ is the probability of no collision.
% The three probabilities correspond to the case that $X_i$ selects the same frequency channel as $X_0$, $X_i$ selects a frequency channel that is adjacent to the one chosen by $X_0$, or $X_i$ selects a distant frequency channel (more than two channels from the one chosen by $X_0$).
ACI can be neglected by setting $\psi=1$ and $K_s=0$, in which case the results
specialize to those presented in \cite{valenti:2012}.
%, in which case $I_i=1$ when $X_i$ selects the same frequency as $X_0$, and $I_i=0$ otherwise.  In this case, it follows that $I_i$ is Bernoulli with probability $P[I_i=1]=p_i$.

% This is a {\em physical interference model} \cite{cardieri:2010}.

Substituting (\ref{eqn:power})  into (\ref{Equation:SINR1}) and dividing the numerator and denominator by $d_0^\alpha P_0$, the SINR is
%   \gamma
%   & = &
%   \frac{ P_0 g_0 10^{\xi_0/10} }{ \displaystyle {\mathcal N} + \sum_{i=1}^M P_i I_i g_i 10^{\xi_i/10} ||X_i||^{-\alpha}}
%   \nonumber \\
%   & = &
\begin{eqnarray}
   \gamma
   & = &
   \frac{ \psi g_0 \Omega_0  }{ \displaystyle \Gamma^{-1} + \sum_{i=1}^M I_i g_i \Omega_i }
   \label{Equation:SINR2}
\end{eqnarray}
where $\Gamma = d_0^\alpha P_{0}/\mathcal{N}$ is the signal-to-noise ratio (SNR)
when the transmitter is at unit distance and the fading and shadowing are absent,
and $\Omega_i = (P_i/P_0)10^{\xi_i/10} |X_i|^{-\alpha}$ is the normalized power of $X_i$. %and $\Omega_0 = 10^{\xi_0/10}||X_0||^{-\alpha}$
%%where $\Gamma = d_0^\alpha P_{0}/\mathcal{N}$ is the signal-to-noise
%%ratio (SNR)  when the transmitter is at unit distance and fading and
%%shadowing are absent,  and
%%\begin{equation}
%%\Omega_{i}=%
%%\begin{cases}
%%10^{\xi_{0}/10}||X_{0}||^{-\alpha} & i=0 \\
%%\displaystyle\frac{P_{i}}{P_{0}}10^{\xi_{i}/10}||X_{i}||^{-\alpha} & i\geq1%
%%\end{cases}
%%\label{eqn:omega}
%%\end{equation}
%is the normalized received power due to $X_{i}$.

\begin{figure*}[b!]
\setcounter{equation}{18}
\hrulefill
\begin{eqnarray}
\mathbb E_{\Omega_i} \left\{ G_{\ell_i} ( \Omega_i ) \right\}
& = &
p_\mathsf{n} \delta_{\ell_i}
+ \frac{ 2 m_i^{m_i} \pi c_i^{\frac{2}{\alpha}} \Gamma(\ell_{i}+m_{i}) }{ \alpha (\ell_{i}! ) \Gamma(m_{i}) \beta_0^{(m_i+\ell_i)} A}
\left \{ p_\mathsf{c}
\left[ I \left( \frac{ \psi c_i}{ r_\mathsf{net}^{\alpha} }   \right)- I \left( \frac{ \psi c_i}{ r_\mathsf{ex}^{\alpha} }  \right) \right] \right.
\left. +
  p_\mathsf{a}
\left[I \left( \frac{ K_s c_i}{ r_\mathsf{net}^{\alpha} }  \right)- I \left( \frac{ K_s c_i}{ r_\mathsf{ex}^{\alpha}} \right)\right]  \right \}.
\label{cdfwithintegral}
\end{eqnarray}
\hrulefill
\setcounter{equation}{23}
\begin{eqnarray}
\bar{F}_{{\mathsf Z}}(z) \hspace{-0.25cm} &=&
\hspace{-0.25cm}
\sum_{s=0}^{m_0-1}
\hspace{-0.10cm}
 \sum_{t=0}^s
\frac{ z^{-t} }{ (s-t)! } \hspace{-0.35cm} \mathop{ \sum_{\ell_i \geq 0}}_{\sum_{i=0}^{M}\ell_i=t}\hspace{-0.35cm}  \int_{0 }^{\infty } \exp \left\{-\frac{\beta m_0  z}{\psi y} \right\} {\left( \frac{\beta m_0 z}{\psi y} \right)}^s \prod_{i=1}^M \left[p_n\delta_{\ell_i}  +  	p_c \Phi_i(y,\psi) + p_a \Phi_i(y,K_s)
\right]  f_{\Omega_0}(y) d y.
\label{shad05}
\end{eqnarray}
\end{figure*}
\setcounter{equation}{5}

\section{Outage Probability} \label{Section:Outage}
\label{Section:OutageProbability}

\subsection{Conditional Outage Probability}

Let $\beta$ denote the minimum SINR required for reliable reception and $\boldsymbol{\Omega }=\{\Omega_{0},...,\Omega _{M}\}$ represent the set of normalized received powers.  An \emph{outage} occurs when the SINR falls below $\beta$.  When conditioned on $\boldsymbol{\Omega }$, the outage probability is
\begin{eqnarray}
   \epsilon( \boldsymbol{\Omega } )
   & = &
   P \left[ \gamma \leq \beta \big| \boldsymbol \Omega \right]\hspace{-0.1cm}.
   \label{Equation:Outage1}
\end{eqnarray}
% Because it is conditioned on $\boldsymbol{\Omega }$, the outage probability depends on the locations of the mobiles and the shadowing factors, which have dynamics over timescales that are much slower than the fading.
By defining a variable
\vspace{-0.1cm}
\begin{eqnarray}
 \mathsf Z & = & \beta^{-1} g_0 \Omega_0 - \sum_{i=1}^M I_i g_i \Omega_i \label{eqn:z}
\end{eqnarray}
the conditional outage probability may be expressed as
\begin{eqnarray}
  \epsilon( \boldsymbol{\Omega } )
  & = &
  P
  \left[
   \mathsf Z \leq \Gamma^{-1} \big| \boldsymbol \Omega
  \right]
  = F_{\mathsf Z} \left( \Gamma^{-1} \big| \boldsymbol \Omega \right) \label{Equation:OutageCDF}
\end{eqnarray}
which is the cumulative distribution function (cdf) of $\mathsf Z$ conditioned on $\boldsymbol \Omega$ and evaluated at $\Gamma^{-1}$.

Let $\bar{F}_{\mathsf Z}\left( z \big| \boldsymbol \Omega \right) = 1 - F_{\mathsf Z}\left( z \big| \boldsymbol \Omega \right)$ denote the complementary cdf of $\mathsf{Z}$ conditioned on $\boldsymbol{\Omega}$.  Assuming that $m_0$ is integer-valued and that signals fade independently, \cite{torrieri:2012} shows that
\begin{eqnarray}
\hspace{-0.5cm}
\bar{F}_{\mathsf Z}\left( z \big| \boldsymbol \Omega \right)
& \hspace{-0.2cm} =  \hspace{-0.2cm} &
e^{-\beta_0 z }
\sum_{j=0}^{m_0-1} {\left( \beta_0 z \right)}^j
 \sum_{k=0}^j
\frac{ z^{-k} }{ (j-k)! }  H_k ( \boldsymbol \Omega )
\label{Equation:NakagamiConditional}
\end{eqnarray}
where  $\beta_0= \beta m_0 /(\psi \Omega_0)$,
\begin{eqnarray}
   H_k ( \boldsymbol \Omega )
   & = &
   \mathop{ \sum_{\ell_i \geq 0}}_{\sum_{i=0}^{M}\ell_i=k}
   \prod_{i=1}^M
   G_{\ell_i} ( \Omega_i ) \label{Equation:Hfunc}
\end{eqnarray}
 the summation in (\ref{Equation:Hfunc}) is over all sets of indices that sum to $k$, and
\begin{eqnarray}
  G_{\ell_i}( \Omega_i )
  & = &
  \int_0^\infty \frac{ y^{\ell_i} }{\ell_i !} e^{-\beta_0 y } f_{\mathsf Y_i}(y) d y
\label{Equation:Integral}
\end{eqnarray}
where $f_{\mathsf Y_i}(y)$ is the pdf of ${\mathsf Y_i} = I_i g_i \Omega_i$. Taking into account the Nakagami fading and the statistics of $I_i$, the pdf of $Y_i$ is
\begin{eqnarray}
  f_{\mathsf Y_i}(y)
  & = &
  p_n \delta(y)
  +  \frac{y^{m_i-1}}{\Gamma(m_i)} \left[
 p_\mathsf{c}
  \left( \frac{m_i}{\psi \Omega_i} \right)^{m_i}
  e^{-\frac{y m_i}{\psi \Omega_i}} \right. \nonumber \\ & &
  \hspace{-1cm}
  \left.  + \; p_\mathsf{a}
  \left( \frac{m_i}{K_s \Omega_i} \right)^{m_i}
  e^{-\frac{y m_i}{K_s \Omega_i}}
  \right]  u(y)
  \label{pdfy}
\end{eqnarray}
% e^{-\frac{y m_i}{\psi \Omega_i}}
% }
where $u(y)$ is the unit-step function and $\delta(y)$ is the Dirac delta function.   By substituting (\ref{pdfy}) into (\ref{Equation:Integral}) and evaluating the integral, we obtain
\begin{eqnarray}
  \hspace{-0.2cm}
  G_{\ell_i}( \Omega_i )
  \hspace{-0.3cm}
  & = &
  \hspace{-0.3cm}
p_\mathsf{n} \delta_{\ell_i}
  \hspace{-0.1cm}
  +
   \hspace{-0.1cm}
    \frac{ \Gamma(\ell_i+m_i) }{ \ell_i! \Gamma( m_i ) }
\left[ p_\mathsf{c} \phi_i( \psi )
+
  p_\mathsf{a}
  \phi_i( K_s )
 \right]  % \nonumber \\
 \label{Equation:Gfunc}
\end{eqnarray}
where $\delta_{\ell} $ is the Kronecker delta function,
% equal to 1 when $\ell=0$, and zero otherwise,
and
\begin{eqnarray}
\hspace{-0.5cm}
  \phi_i( x )
  & = &
  \left( \frac{x \Omega_i}{ m_i} \right)^{\ell_i}
  \left(
   \frac{ x \beta_0 \Omega_i}{m_i} + 1
 \right)^{-(m_i+\ell_i)}\hspace{-1.25cm}.
\end{eqnarray}

%The conditional outage probability may be calculated once the location of the mobiles and the shadowing factors are specified. Thus, performance can be calculated for any snapshot of the network topology. Insight into the average network performance within a geographical region may be obtained by computing spatial averages, as described next.

\subsection{Spatially Averaged Outage Probability}

Assume that a fixed number $M$ of interfering mobiles are independently and uniformly distributed over the region of area $A$; i.e., that the interfering mobiles are drawn from a {\em binomial point process} (BPP) of intensity $\lambda=M/A$ \cite{stoyan:1996}.
%Because it is conditioned on ${\boldsymbol \Omega}$, the conditional outage probability $\epsilon( \boldsymbol{\Omega } )$ given in (\ref{Equation:OutageCDF}) depends on the location of the mobiles and the values of the shadowing factors.
Let $\epsilon(\lambda)$ denote the corresponding spatially averaged outage probability, which is found by taking the expectation of ${F}_{\mathsf Z}(\Gamma^{-1}|\boldsymbol \Omega)$ with respect to ${\boldsymbol \Omega}$:
\begin{eqnarray}
   \epsilon(\lambda)
   &=&
   \mathbb E_{\boldsymbol \Omega} \left\{
   {F}_{\mathsf Z}\left( \Gamma^{-1} \big| \boldsymbol \Omega \right)
   \right\}
   =
   {F}_{\mathsf Z}\left( \Gamma^{-1} \right). \label{expectation}
\end{eqnarray}

First, consider a system that has no shadowing ($\xi_i = 0$, for all $i$) and a fixed location $X_0$ so that $\Omega_0$ and $\beta_0$ are constants.  By assuming that the $\{\Omega_i\}, i > 0,$ are independent,
taking the expectation with respect to $\boldsymbol \Omega$ allows the complementary cdf of $\mathsf{Z}$ to be expressed as
\begin{eqnarray}
\hspace{-0.5cm}
\bar{F}_{\mathsf Z}\left( z \right)
& \hspace{-0.3cm} =  \hspace{-0.3cm} &
e^{-\beta_0 z }
\sum_{j=0}^{m_0-1} {\left( \beta_0 z \right)}^j
 \sum_{k=0}^j
\frac{ z^{-k} }{ (j-k)! }  \mathbb E_{\boldsymbol \Omega} \left\{ H_k ( \boldsymbol \Omega ) \right\}
\label{Equation:NakagamiConditionalAvg}
\end{eqnarray}
where
\begin{eqnarray}
   \mathbb E_{\boldsymbol \Omega} \left\{ H_k ( \boldsymbol \Omega ) \right\}
   & = &
   \mathop{ \sum_{\ell_i \geq 0}}_{\sum_{i=0}^{M}\ell_i=k}
   \prod_{i=1}^M
   \mathbb E_{\Omega_i} \left\{ G_{\ell_i} ( \Omega_i ) \right\}.  \label{Equation:HfuncAvg}
\end{eqnarray}
Since the $\{\Omega_i\}$, $i = 1, ..., M$, are uniformly distributed on the annulus, they each have pdf
\begin{eqnarray}
   f_{\Omega_i}( \omega )
     \hspace{-0.2cm}
   & = &
     \hspace{-0.2cm}
      \frac{2\pi}{A} c_i^{2/\alpha}  \omega^{-\left( \frac{2+\alpha}{\alpha} \right) }
\hspace{0.4cm}
      \mbox{for $\frac{c_i}{r_\mathsf{net}^{\alpha}} \leq \omega \leq \frac{c_i}{r_\mathsf{ex}^{\alpha}}$}  \label{pdf_mho2}
\end{eqnarray}
and zero elsewhere, where $c_i = (P_i/P_0)$. Using (\ref{pdf_mho2}), the expectation in (\ref{Equation:HfuncAvg}) evaluates to
(\ref{cdfwithintegral}) at the bottom of the page,  where
\setcounter{equation}{19}
\vspace{-0.25cm}
\small
\begin{eqnarray}
 I( x )
=
 \frac{  {_2F_1 \left( \left[ m_i + \ell_i, m_i + \frac{2}{\alpha} \right]; m_i+\frac{2}{\alpha}+1;-\frac{m_i }{x \beta_0} \right)} }{ x^{m_i+\frac{2}{\alpha}} \left( m_i + \frac{2}{\alpha} \right) }
\label{Ihyp}
\end{eqnarray}
\normalsize
and $_2F_1([a,b],c,z)$ is the Gauss hypergeometric function.

\subsection{Log-Normal Shadowing}
In the presence of log-normal shadowing, the $\{ \xi_i \}$ are independent and identically distributed zero-mean Gaussian with standard deviation $\sigma_\mathsf{s}$ dB.
When $X_0$ is fixed but shadowing is present, $\Omega_0$ becomes a log-normal random variable with pdf
\begin{eqnarray}
f_{\Omega_0}(\omega)
=
\frac{10 \left(2 \pi \sigma_\mathsf{s}^2\right)^{-\frac{1}{2}}}{\ln(10) \omega}  \exp\left \{ {-\frac{10^2 \log_{10}^2 \left( ||X_0||^{\alpha} \omega \right) }{2 \sigma_\mathsf{s}^2}}\right \}
\label{shad02}
\end{eqnarray}
for $0 \leq \omega \leq \infty$, and zero elsewhere, while $\Omega_i$, $i = 1, ..., M$, has pdf \cite{valenti:2012}
\vspace{-0.15cm}
\begin{eqnarray}
f_{\Omega_i}(\omega)
\hspace{-0.3cm}
& = &
\hspace{-0.3cm}
\frac{\pi c_i^{2/\alpha} }{ A \alpha }
\left[ \zeta \left(c_i \omega r_{\mathsf{net}}^{\alpha} \right) -\zeta \left( c_i \omega r_{\mathsf{ex}}^{\alpha} \right) \right]
\omega^{-\left( \frac{2+\alpha}{\alpha} \right) }
\label{shad8}
\end{eqnarray}
for $0 \leq \omega \leq \infty$, and zero elsewhere,
where
\begin{eqnarray}
\zeta( z )
\hspace{-0.2cm}
& = &
\hspace{-0.2cm}
 \mbox{erf} \left(  \frac{\sigma_\mathsf{s}^2 \ln^2(10)-50 \alpha \ln\left( z \right)}{5 \sqrt{2} \alpha \sigma_\mathsf{s} \ln(10)}\right) \exp \left\{ \frac{\sigma_\mathsf{s}^2 \ln^2(10)}{50 \alpha^2} \right\}. \nonumber \\
\label{zeta}
\end{eqnarray}
%while
%\begin{eqnarray}
%f_{\Omega_0}(\omega)
%=
%\frac{10 \left(2 \pi \sigma_s^2\right)^{-\frac{1}{2}}}{\ln(10) \omega}  \exp\left \{ {-\frac{10^2 \log_{10}^2 \left( ||X_0||^{\alpha} \omega \right) }{2 \sigma_s^2}}\right \}
%\label{shad02}
%\end{eqnarray}
%for $0 \leq \omega \leq \infty$, and zero elsewhere.

Using the definition of $\beta_0$ and the pdfs of the $\{\Omega_i\}$, taking the expectation of $\bar{F}_{{\mathsf Z}}(z | \boldsymbol \Omega)$ with respect to $\Omega$ allows the complementary cdf of $\mathsf{Z}$ to be expressed as
(\ref{shad05}) at the bottom of the previous page,  where
\setcounter{equation}{24}
\begin{eqnarray}
\Phi_i(y,\chi)
\hspace{-0.2cm}
& = & \nonumber \\
& &
\hspace{-2cm}
\frac{
\Gamma(\ell_{i}+m_{i})
}{
\ell_{i}!
\Gamma(m_{i})
}
\int_{0 }^{\infty }
f_{\Omega_i} ( \omega )
\left( \frac{\chi \omega}{ m_i } \right)^{\ell_i}
\left(
\frac{ \chi \beta m_0 \omega }{ \psi m_i y } + 1
\right)^{-(m_i + \ell_i)} \hspace{-0.2cm}
  d \omega.
  \nonumber \\
\label{Psi_function}
\end{eqnarray}
The integral in (\ref{Psi_function}) can be evaluated numerically by Simpson's method, which provides a good tradeoff between accuracy and speed, and then the integral in  (\ref{shad05}) can be evaluated through Monte Carlo simulation.

\section{Modulation-Constrained Transmission Capacity}\label{Section:TC}
The spatially averaged outage probability $\epsilon( \lambda )$ is often constrained to not exceed a maximum value $\zeta$ $\in$ $\left[ 0,1 \right]$; i.e., $\epsilon(\lambda) \leq \zeta$.  Under such a constraint, the maximum density of transmissions is of interest, which is quantified by the {\em transmission capacity} (TC) \cite{weber:2005}.  The TC represents the spatial spectral efficiency; i.e., the rate of successful data transmission per unit area.
%Under outage constraint $\zeta$, the TC in units of bits-per-second (bps) per unit area is
%\begin{eqnarray}
%\tau\left(\zeta \right)
% & = &
%\epsilon^{-1}(\zeta)(1-\zeta) b
%\label{TC_definition}
%\end{eqnarray}
%where $b$ is the per-link throughput in the absence of an outage, in units of bps, $\epsilon^{-1}(\zeta)$ is the density of the underlying process whose spatially averaged outage probability satisfies the constraint $\epsilon(\lambda) \leq \zeta$ with equality,
%\footnote{Since $\epsilon$ is a monotonically increasing function of $\lambda$, the TC is maximized when the constraint $\epsilon \leq \zeta$ is met with equality.},
%and $(1-\zeta)$ ensures that only successful transmissions are counted.
% With appropriately normalized variables, the TC can assume units of bits-per-second per Hz per $m^2$ (bps/Hz/$m^2$). is a measure of the spatial intensity of transmissions and has the units of number of (successful) transmissions per unit area.
%The TC embodies the tradeoff between the outage probability $\epsilon$ and spatial density $\lambda$.
%In (\ref{TC_definition}), the tradeoff between  outage probability and spatial density
%is quantified by fixing $\epsilon=\zeta$ and varying $\lambda$.
Alternatively, the spatially averaged TC can be found by fixing $\lambda$ and allowing $\epsilon$ to vary, in which case it is expressed as
% An alternative representation
% which is more applicable to the approach taken in this paper,
% is to fix $\lambda$ and allow $\epsilon$ to vary.  This allows the TC to be expressed as a % function of $\lambda$:
\begin{eqnarray}
\tau\left( \lambda \right)
 & = &
\lambda (1 - \epsilon(\lambda) )b.
\label{TC_lambda}
\end{eqnarray}
%where $\epsilon$ is the spatially averaged outage probability when the underlying point process has intensity $\lambda$.

As originally defined in \cite{weber:2005}, the transmission capacity is a function of the SINR threshold $\beta$ and is found without making any assumptions about the type of modulation or channel code.  In practice, $\beta$ is a function of the modulation and coding that is used.   Let $C( \gamma )$ denote the maximum achievable rate, in bps, that can be supported by the chosen modulation at an instantaneous SINR  $\gamma$ assuming equally likely input symbols; i.e., it is the modulation-constrained capacity or symmetric-information rate.  If a capacity-achieving rate-$R$ code is used, then an outage will occur when $C(\gamma) \leq R$.   Since $C(\gamma)$ is monotonic, it follows that $\beta$ is the value for which $C(\beta)=R$, and therefore we can write $\beta = C^{-1}(R)$.  In a block-fading channel, the outage probability with SINR threshold $\beta = C^{-1}(R)$ provides an accurate prediction of the codeword error rate \cite{torrieri:2011}.
%\begin{figure}[t]
%\centering
%\vspace{0.1cm}
%\includegraphics[width=9cm]{figures/FigD}
%\vspace{-0.5cm}
%\caption{ The symmetric-information rate $C(h,\gamma)$ of noncoherent binary CPFSK with modulation indices $h=[0.2,0.4,0.6,1]$.  The markers were found through Monte Carlo integration while the curves are a polynomial fit. \label{Figure:InfoRate} }
%\vspace{-0.5cm}
%\end{figure}
% Frequency-hopping systems often use noncoherent CPFSK \cite{torrieri:2011,cheng:ciss2007}.  A key parameter in CPFSK systems is the {\em modulation index} $h$, which represents the relative spacing between tones. To emphasize the dependence of the capacity on $h$, we use $C(h,\gamma)$ to denote the symmetric-information rate of CPFSK with modulation index $h$.
The symmetric-information rate of noncoherent CPFSK is given in \cite{torrieri:2008} for various $h$, and is found by computing the average mutual information between the input and the output of the noncoherent AWGN channel.
% In a noncoherent system, the mutual integration integral cannot be solved in closed form and is most effectively found using Monte Carlo integration.
% Fig. \ref{Figure:InfoRate}, which is adapted from Fig. 1 of \cite{cheng:ciss2007}, shows the symmetric information rate of binary CPFSK as a function of $\gamma$ for various $h$.  In the figure, Monte Carlo integration is used to obtain the symmetric information rate at discrete values of $\gamma$, and a polynomial fit is used to determine the other values.    For any value of $h$, the value of the SINR threshold $\beta$ can be found from the corresponding curve by finding the value of $\gamma$ for which $C(h,\gamma)=R$.
As an example, when $R=1/2$ and $h=1$, the required $\beta = 3.7$ dB.
%, it was found that in practice, and over a wide range of code rates, turbo-coded noncoherent CPFSK is consistently about 1 dB away from the corresponding modulation-constrained capacity limit.  Thus, the $\beta$ required in practice will generally be higher than the value obtained from the capacity limit by a small margin.  For instance, if the actual system requires a 1 dB margin over the capacity limit, then the SINR threshold for noncoherent binary CPFSK with $R=1/2$ and $h=1$ should be set to $\beta = 4.7$ dB.

The maximum data transmission rate is determined by the bandwidth $B/L$ of a frequency channel, the fractional in-band power $\psi$, the spectral efficiency of the modulation, and the code rate.  Let $\eta$ be the spectral efficiency of the modulation, given in symbols per second per Hz, and defined by the symbol rate divided by the $100\psi$\%-power bandwidth of the modulation.  Since we assume many symbols per hop, the spectral efficiency of CPFSK can be found by numerically integrating (3.4-61) of \cite{proakis:2008} and then inverting the result.  To emphasize the dependence of $\eta$ on $h$ and $\psi$, we denote the spectral efficiency of CPFSK as $\eta(h,\psi)$ in the sequel.  When combined with a rate-$R$ code, the spectral efficiency becomes $R \eta(h,\psi)$ bps per Hertz (bps/Hz), where $R$ is the ratio of information bits to code symbols.  Since the signal occupies a frequency channel with $100\psi$\%-power bandwidth $B/L$ Hz, the maximum data rate supported by a single link operating with a duty factor $D$ is
\begin{eqnarray}
b
& = &
\frac{ R D \eta(h,\psi) B}{L}
\label{throughput}
\end{eqnarray}
bps.  Substituting (\ref{throughput}) into (\ref{TC_lambda}) and dividing by the system bandwidth $B$  gives the normalized and spatially averaged {\em modulation-constrained} transmission capacity (MCTC),
\begin{eqnarray}
  \tau'(\lambda)
  & = &
  \frac{\lambda R D \eta(h,\psi) (1-\epsilon(\lambda)) }{L} \label{Equation:TCnorm}
\end{eqnarray}
which assumes units of bps/Hz per unit area.  In contrast with (\ref{TC_lambda}), this form of transmission capacity explicitly takes into account the code rate $R$, the spectral efficiency of the modulation $\eta(h,\psi)$, and the number of frequency channels.

%Note that (\ref{Equation:TC}) requires that the outage probability be independent of the location of the reference receiver, which is true when the network is assumed to be infinite and uniform.  Otherwise, the throughput needs to be computed at each possible location, and the spatial average taken.

\section{Network Optimization}\label{Section:Optimization}
Let $\boldsymbol \theta = (L,R,h,\psi)$ and $C(\boldsymbol \theta)$ represent a cost function.
The goal of the network optimization is to determine the $\boldsymbol \theta$ that minimizes an appropriate cost function.   Fix the value of $\lambda$, and let $\tau'( \boldsymbol \theta )$ represent the normalized MCTC at that $\lambda$ as a function of $\boldsymbol \theta$.  We chose as our cost function $C( \boldsymbol \theta ) = - \tau'( \boldsymbol \theta )$ so that the optimization maximizes the normalized MCTC.
%We begin in Subsection \ref{Section:Exhaustive} by exhaustively evaluating the MCTC for a wide range of discretized $\boldsymbol \theta$ for a representative set of channel conditions.  The results suggest that the MCTC is convex.  Then, in Subsection \ref{Section:Search} we present an unconstrained nonlinear optimization technique based on the Nelder-Mead downhill simplex method  described in \cite{Lagarias:1998}, which provides an efficient solution to the optimization problem.  Results of the optimization are given in Subsection \ref{Section:Results}.

% Because the search-space is a four-dimensional region and the cost function is nonlinear, finding an efficient solution to the optimization problem is challenging.

\subsection{Exhaustive Evaluation}\label{Section:Exhaustive}
As an example, consider a network with radius $r_{\mathsf{net}} = 2$, an exclusion zone of radius $r_{\mathsf{ex}} = 0.25$, and
$M=50$ potentially interfering mobiles drawn from a BPP with density $\lambda = 50/(\pi(2^2-0.25^2)) \approx 4$. The duty factor is $D=1$,  the source is located at distance $|X_0|=1$, the path-loss exponent is $\alpha = 3$, the SNR is $\Gamma = 10$ dB, and the shadowing variance is $\sigma_\mathsf{s} = 8$ dB.  A {\em mixed-fading} model is considered, with $m_0 = 4$ and $m_i = 1$ for $i \geq 1$.  Mixed fading characterizes a typical situation where the source transmitter is in the line-of-sight to the receiver, but the interfering mobiles are not in the line-of-sight.  Three degrees of spectral splatter are considered: No spectral splatter, minimal spectral splatter ($\psi = 0.99$), and moderate spectral splatter ($\psi = 0.96$).   When spectral splatter is neglected, $\eta$ is computed using the $99$-percent power bandwidth of CPFSK, but $p_\mathsf{a}$ is set to zero.  When spectral splatter is considered, $\eta$ is computed using the $100\psi$-percent power bandwidth, and $p_\mathsf{a}$ is set according to (\ref{eqn:p_a}).   For each spectral-splatter case, the MCTC was computed for a wide range of discretized $\boldsymbol \theta$ that include all integer $L \in [1,200]$, all $R \in (0,1)$ in increments of $0.01$, and all $h \in (0,1]$ in increments of $0.01$.

\begin{figure}[t]
\centering
\hspace{-0.5cm}
\includegraphics[width=9cm]{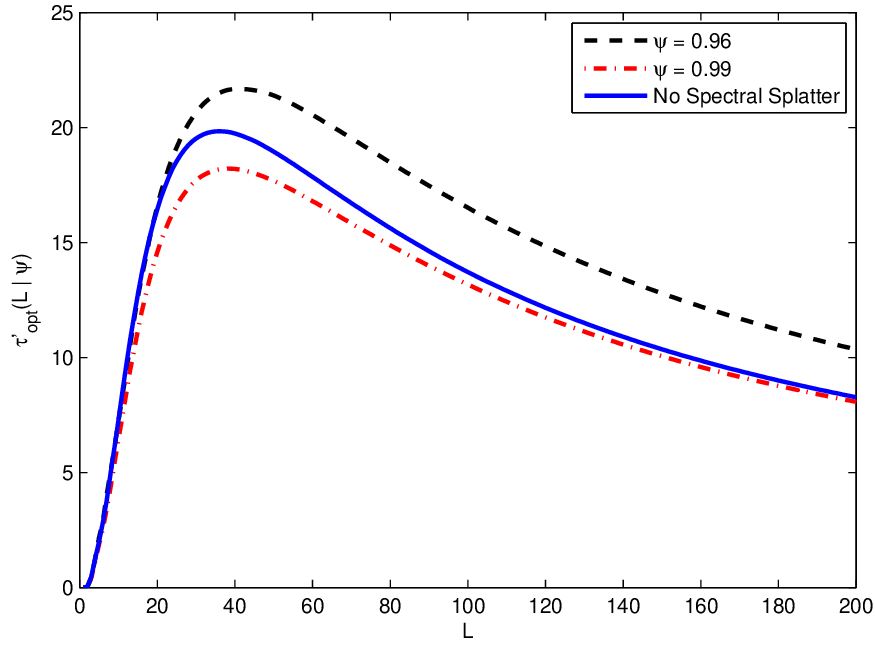}
\vspace{-0.3cm}
\caption{The MCTC $\tau_{\mathsf{opt}}'(L | \psi)$ as a function of the number of frequency channels $L$.  For each value of $L$, the modulation-index $h$ and code rate $R$ are varied to maximize the TC.   Top curve: $\psi$ = 0.96.  Middle curve: ACI due to spectral splatter is neglected.  Bottom curve: $\psi$ = 0.99.
\label{Figure:L} }
\vspace{-0.65cm}
\end{figure}

Because $\tau'( \boldsymbol \theta)$ has a four-dimensional argument, it is not easily visualized.  Fig. \ref{Figure:L} represents the function by fixing $\psi$, varying $L$, and  maximizing the value of $\tau'( \boldsymbol \theta)$ over $R$ and $h$.  To emphasize the dependence on only one parameter and optimization over the other parameters, the function is written as a function of only that variable
\begin{eqnarray}
   \tau_{\mathsf{opt}}'( L | \psi )
   & = &
   \max_{R,h} \tau' (\boldsymbol \theta ).
\end{eqnarray}

Similar figures found by fixing each of $R$ and $h$ while optimizing over the other parameters suggest that $C(\boldsymbol \theta)$ has global minima over all feasible $\boldsymbol \theta$, which is further confirmed by an inspection of the complete multidimensional surface.  It follows that the optimal values of each parameter can be found by locating the peaks in the corresponding figure.  Fig.  \ref{Figure:L}  shows that, for $\psi=0.99$, performance degrades when spectral splatter is taken into account.  However, when the amount of spectral splatter is increased (by decreasing $\psi$ to 0.96), the performance improves.  This illustrates the potential gain that can be achieved by
jointly optimizing over $\psi$ and the other parameters.

%
%\begin{figure}[t]
%\centering
%\hspace{-0.5cm}
%\includegraphics[width=9cm]{figures/Fig3}
%\vspace{-0.5cm}
%\caption{The MCTC $\tau_{\mathsf{opt}}'(R)$ as a function of the code rate $R$.  For each value of $R$, the modulation-index $h$ and number of frequency channels $L$ are varied to maximize the MCTC.   Top curve: $\psi$ = 0.96, which is the optimal value of fractional-in-band power.  Middle curve: ACI due to spectral splatter is neglected.  Bottom curve: $\psi$ = 0.99.
%\label{Figure:R} }
%\vspace{-0.5cm}
%\end{figure}
%
%\begin{figure}[t]
%\centering
%\hspace{-0.5cm}
%\includegraphics[width=9cm]{figures/Fig4}
%\vspace{-0.5cm}
%\caption{The MCTC $\tau_{\mathsf{opt}}'(h)$ as a function of the modulation-index $h$.  For each value of $h$, the code rate $R$ and number of frequency channels $L$ are varied to maximize the TC.   Top curve: $\psi$ = 0.96, which is the optimal value of fractional-in-band power.  Middle curve: ACI due to spectral splatter is neglected.  Bottom curve: $\psi$ = 0.99.
%\label{Figure:h} }
%\vspace{-0.5cm}
%\end{figure}

\subsection{Downhill Simplex Optimization}\label{Section:Search}
The
%While
exhaustive evaluation of $\tau'( \boldsymbol \theta )$ over a large set of discretized $\boldsymbol \theta$
%identifies the optimal $\boldsymbol \theta$ to within the precision of the discretization, computing the MCTC for such a large number of points
is computationally intensive, especially when the Nakagami factor $m_0$ is large.  This motivates the use of a more efficient search technique.  Because $\tau'( \boldsymbol \theta )$ is a complicated nonlinear function of $\boldsymbol \theta$, it is difficult to obtain analytical or numerical derivatives.  For this reason, a direct search is preferred over a gradient search.  The Nelder-Mead method of downhill simplex optimization \cite{Lagarias:1998} is an appropriate and efficient solution for this optimization problem.

In four dimensions, the Nelder-Mead method works by evaluating the cost function $C(\boldsymbol \theta)$ at the five corners of a pentachoron; i.e., a 4-dimensional convex regular polytope or hyperpyramid.  After each iteration, one corner of the pentachoron is moved until it contains the minimum, at which point the pentachoron is made smaller.   In our implementation, the first corner  is initially at $(L_\mathsf{e},R,h, \psi) = (20, 0.5, 0.5, 0.975)$ and the other corners are at distances $1$, $0.025$, $0.025$, and $0.005$ from the first corner along each of the four dimensions.  Although $L$ needs to be an integer, during the optimization we allow it to be real valued to ensure a continuous optimization surface.

Let $\boldsymbol \theta_1, ...,  \boldsymbol \theta_5$ represent the corners of the pentachoron sorted in ascending cost; i.e., $C(\boldsymbol \theta_1) \leq C(\boldsymbol \theta_2) \leq ... \leq C(\boldsymbol \theta_5)$.  An iteration proceeds by first {\em reflecting} $\boldsymbol \theta_5$ across the opposing face of the pentachoron to produce a candidate corner $\boldsymbol \theta_{\mathsf r}$ whose cost is computed.  If $C( \boldsymbol \theta_1 ) \leq C(\boldsymbol \theta_{\mathsf r}) < C(\boldsymbol \theta_4)$, then  corner $\boldsymbol \theta_5$ is replaced by $\boldsymbol \theta_{\mathsf r}$.  The points are re-sorted and the algorithm moves on to the next iteration.  If
$C(\boldsymbol \theta_{\mathsf r}) < C(\boldsymbol \theta_1)$, then an {\em expanded} pentachoron is considered by doubling the distance from $\boldsymbol \theta_{\mathsf r}$ to the face defined by the other four corners, thereby producing another candidate corner $\boldsymbol \theta_{\mathsf s}$ whose cost is computed.  If $C(\boldsymbol \theta_{\mathsf s}) < C(\boldsymbol \theta_{\mathsf r})$, then the expanded pentachoron is accepted (by replacing $\boldsymbol \theta_5$ with $\boldsymbol \theta_{\mathsf s}$), otherwise the reflected (but unexpanded) pentachoron is accepted by replacing $\boldsymbol \theta_5$ with $\boldsymbol \theta_{\mathsf r}$.  If $C(\boldsymbol \theta_{\mathsf r}) \geq C(\boldsymbol \theta_4)$ then a {\em contraction} is performed by halving the distance between the better of $\boldsymbol \theta_5$ and $\boldsymbol \theta_{\mathsf r}$ and the face defined by the other four corners.  The contracted pentachoron is accepted if this new corner has a lower cost than the one it displaced.  Otherwise, if none of the above conditions is satisfied then the minimum must lie inside the pentachoron, so it is {\em shrunk} by halving the length of each edge while maintaining the same centroid.

\subsection{Optimization Results}\label{Section:Results}

By using the Nelder-Mead method, optimization results were obtained for a range of densities $\lambda$ with interfering mobiles drawn from a BPP.  As with the example used to generate Fig. \ref{Figure:L},
% to \ref{Figure:h},
the number of interfering mobiles is $M=50$, the duty factor is $D=1$, the source is located at distance $|X_0|=1$, the path-loss exponent is $\alpha = 3$, the SNR is $\Gamma = 10$ dB, and the exclusion zone has radius $r_{\mathsf{ex}} = 0.25$.  Three fading models are considered: Rayleigh fading ($m_i = 1$ for all $i$), Nakagami fading ($m_i = 4$ for all $i$), and mixed fading ($m_0 = 4$ and $m_i = 1$ for $i \geq 1$).  Both unshadowed ($\sigma_\mathsf{s} = 0$ dB) and shadowed ($\sigma_\mathsf{s} = 8$ dB) environments are considered.  Two values of  $r_{\mathsf{net}}$ are considered:  $r_{\mathsf{net}} = 2$, and  $r_{\mathsf{net}} = 4$, corresponding to a moderately dense ($\lambda \approx 4$) and sparse ($\lambda \approx 1$) network, respectively.

\begin{figure}[t]
\centering
\includegraphics[width=9cm]{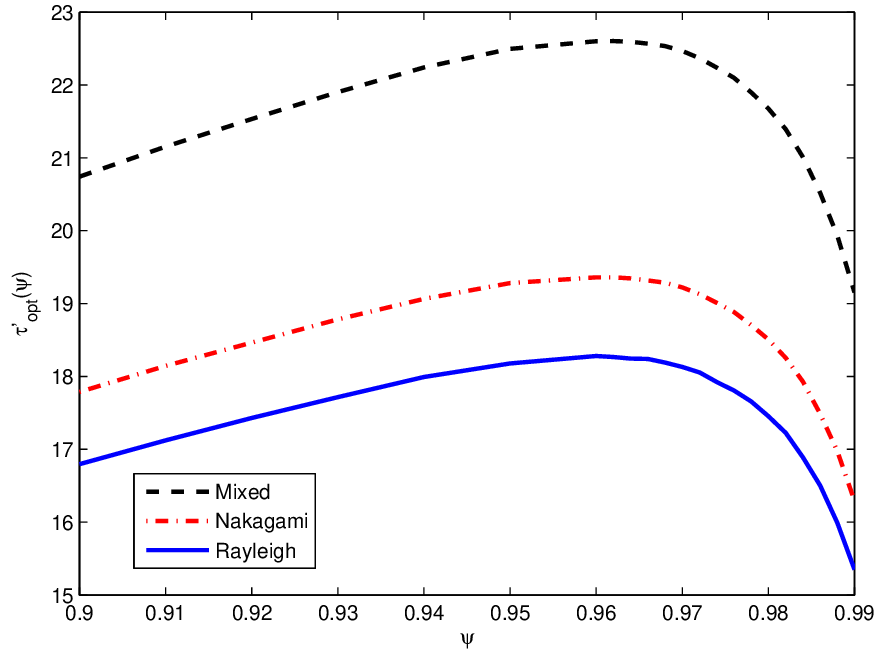}
\vspace{-0.75cm}
\caption{ Maximum normalized transmission capacity $\tau'_{opt}(\psi)$ as a function of the fractional in-band power.  For each value of $\psi$, the modulation-index $h$, the  number of frequency channels $L$, and code rate $R$ are varied to maximize the TC.   Top curve: Mixed fading.  Middle curve: Nakagami Fading.  Bottom curve: Rayleigh Fading.
\vspace{-0.25cm}
\label{Figure:psi} }
\end{figure}

The influence of $\psi$ on the optimal transmission capacity is shown in Fig. \ref{Figure:psi} for the three fading models in the presence of shadowing ($\sigma_\mathsf{s}=8$) with $r_\mathsf{net}=2$.  The curves were drawn by fixing $\psi$ and optimizing over $(L,R,h)$.  From these curves, the optimal value of $\psi$ is identified to be about 0.96 for all three fading models.  As seen in the figures, by choosing $\psi=0.96$, there is an increase in TC relative to the typical, but arbitrary choice of $\psi=0.99$. The resulting optimal parameters when $\psi = 0.96$ are found to be $\{L,R,h\} = \{ 38, 0.64, 0.81 \}$ in the mixed-fading environment.  When the ACI due to spectral splatter is neglected and the $99$-percent power bandwidth used during the optimization, the resulting parameter values are $\{L,R,h\} = \{ 24, 0.68, 0.59 \}$ in the mixed-fading environment \cite{valenti:2012}.  Thus, when ACI is taken into account during the optimization, more frequency channels $L$ are optimal, and the optimal value of modulation index $h$ is larger.

Table \ref{Table:Main} shows the optimal values of $L, R, h$ and $\psi$ for the two network radii, three fading models, and two shadowing variances, along with with the corresponding $\tau'_\mathsf{opt}$.    The optimal fractional in-band powers are $\psi=0.96$ for the $r_\mathsf{net}=2$ network and $\psi=0.95$ for the $r_\mathsf{net}=4$ network. For the Rayleigh channel, shadowing slightly improves performance, but for the Nakagami and mixed-fading channels, shadowing slightly degrades the performance. Increasing the network density (by decreasing $r_\mathsf{net}$) increases the transmission capacity, and requires an increased $L$, $R$, and $\psi$ and a decreased $h$.

 \begin{table}
  \caption{Optimal parameter values and transmission capacity.  \label{Table:Main} }
\centering
 \begin{tabular}{|c|c|c|c|c|c|c|c|c|}
  \hline
  $r_\mathsf{net}$ & $\sigma_\mathsf{s}$ & $m_0$ & $m_i$ & $L$ & $R$ & $h$ & $\psi$  & $\tau'_\mathsf{opt} $  \\
  \hline
  \hline
  2         & 0            &  1    &   1   & 36  & 0.64  & 0.81  & 0.96  &  17.74 \\
  \cline{3-9}
            &              &  4    &   4   &   45 & 0.64 & 0.81  & 0.96  &   19.88 \\
  \cline{3-9}
            &              &  4    &   1   &  40 & 0.64 &  0.81 &  0.96  & 22.59  \\
  \cline{2-9}
            & 8            &  1    &   1   &  34 & 0.64 &  0.81 &  0.96  &  18.29  \\
  \cline{3-9}
            &              &  4    &   4   &  44 & 0.66 & 0.81  & 0.96  & 19.36     \\
  \cline{3-9}
            &              &  4    &   1  & 38 &  0.64 & 0.81 & 0.96  &22.09 \\
  \hline
  \hline
  4         & 0            &  1    &   1   &  13  & 0.57 & 0.85& 0.95  & 11.92 \\
  \cline{3-9}
            &              &  4    &   4   & 16   & 0.54&0.85 & 0.95  &  13.23  \\
  \cline{3-9}
            &              &  4    &   1   & 15    &0.56 & 0.85 &  0.95  & 14.64   \\
  \cline{2-9}
            & 8            &  1    &   1   & 12     &0.58 & 0.85 & 0.95 & 12.22 \\
  \cline{3-9}
            &              &  4    &   4   & 15    & 0.54& 0.85 & 0.95  & 13.13  \\
  \cline{3-9}
            &              &  4    &   1   &  14    & 0.57& 0.85 & 0.95  & 14.55 \\
  \hline
  \end{tabular}
  \vspace{-0.5cm}
\end{table}

\subsection{Effect of Normalized Distance}
\begin{figure}[t]
\centering
\hspace{-0.45cm}
\includegraphics[width=9cm]{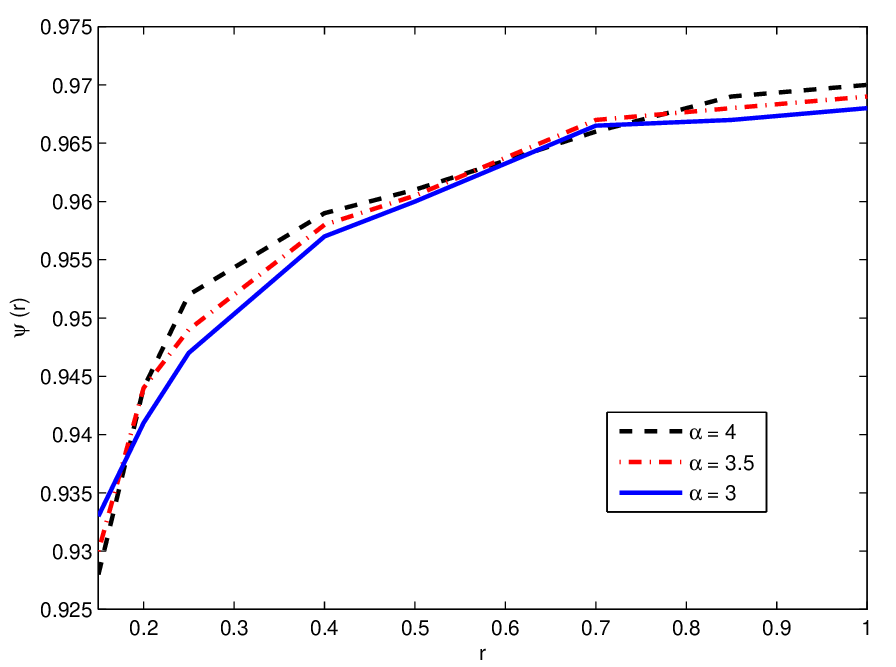}
\vspace{-0.3cm}
\caption{Optimal fractional in-band power $\psi(r)$ as a function of normalized transmitter distance $r=|X_0|/r_\mathsf{net}$.   Results are shown for three different path-loss coefficients. \label{Figure:r} }
\vspace{-0.5cm}
\end{figure}

%As can be seen in Table \ref{Table:Main},
For a fixed $|X_0|$ and $M$, the performance and optimal value of $\psi$ depends on $r_\mathsf{net}$. More generally, the optimal value of $\psi$ can be identified for a given $M$ for any arbitrary normalized distance $r=|X_0|/r_\mathsf{net}$.
 At each $r$, an optimization is performed to determine the optimal $\boldsymbol \theta$ and the corresponding TC.
 The dependence of the optimal $\psi$ on $r$ is shown in Fig. \ref{Figure:r} for three values of path-loss coefficient, $\alpha = \{3, 3.5, 4\}$, using the mixed-fading model and shadowing ($\sigma_\mathsf{s} = 8$ dB).  From the results, it is observed that the optimal $\psi$ increases with increasing separation $r$ and increasing path-loss coefficient $\alpha$.

% \vspace{-0.1cm}

\section{Conclusion} \label{Section:Conclusion}
When used with coded CPFSK, the performance of frequency-hopping ad hoc networks depends on the modulation index, the code rate, the number of frequency channels, and the fractional in-band power $\psi$.  The procedure outlined in this paper enables the optimization of the parameters in the presence of shadowing and Nakagami fading with interfering mobiles drawn from an arbitrary point process.  The proper choice of $\psi$ involves a tradeoff between the transmission rate and the amount of adjacent-channel interference.     For a given symbol rate, $\psi$ can be increased by increasing the frequency separation between adjacent frequency channels.  The result will be decreased adjacent-channel interference, but this comes at the cost of reducing the total number of frequency channels, which results in more frequent co-channel collisions.

\balance

%\vspace{-0.1cm}

\bibliographystyle{ieeetr}
\bibliography{./icc2012refs}

\end{document}